\documentclass[twocolumn]{jpsj3}

\usepackage{times}
\usepackage{graphicx}

\def\runauthor{Takashi {\sc Hotta}}

\title{Construction of a Microscopic Model for Yb and Tm Compounds \\
on the Basis of a $\mib{j}$-$\mib{j}$ Coupling Scheme}

\author{Takashi {\sc Hotta}}

\inst{Department of Physics, Tokyo Metropolitan University,
1-1 Minami-Osawa, Hachioji, Tokyo 192-0397, Japan}

\recdate{\today}

\abst{
We provide a prescription to construct a microscopic model
for heavy lanthanide systems such as Yb and Tm compounds
by exploiting a $j$-$j$ coupling scheme.
Here we consider a situation with a large spin-orbit coupling,
in which $j$=5/2 sextet is fully occupied,
while $j$=7/2 octet is partially occupied,
where $j$ denotes total angular momentum.
We evaluate crystalline electric field potentials and
Coulomb interactions among the states of the $j$=7/2 octet
to construct a local Hamiltonian in the $j$-$j$ coupling scheme.
Then, it is found that the local $f$-electron states composed of
the $j$=7/2 octet agree quite well with those of seven $f$ orbitals
even for a realistic value of the spin-orbit coupling.
As an example of the application of the present model,
we discuss low-temperature multipole states
of Yb- and Tm-based filled skutterudites
by analyzing multipole susceptibility
of the Anderson model in the $j$-$j$ coupling scheme
with the use of a numerical renormalization group technique.
>From the comparison with the numerical results of
the seven-orbital Anderson model,
it is concluded that the multipole state is also well reproduced
by the $j$-$j$ coupling model, even when we include
the hybridization between conduction and $f$ electrons
for the realistic value of the spin-orbit coupling.
Finally, we briefly discuss future applications of the present
prescription for theoretical research on heavy lanthanide compounds.
}

\kword{$j$-$j$ coupling scheme, microscopic model, heavy electrons,
multipole, filled skutterudites}

\begin{document}
\maketitle

%
%
\section{Introduction}

It has been widely recognized that emergence of heavy electron state
is understood from the competition in $f$-electron duality nature,
\cite{Kondo40}
i.e., itinerancy due to Kondo effect \cite{Kondo1,Kondo2a,Kondo2b} vs.
localization due to Ruderman-Kittel-Kasuya-Yosida (RKKY) interaction.
\cite{RKKY1,RKKY2,RKKY3}
There appears a quantum phase transition at zero temperature
in the $f$-electron state between itinerant and localized regions.
In such a competing region, we have frequently observed
unconventional superconductivity and non-Fermi liquid behavior
due to the effect of quantum critical fluctuations.
\cite{review1,review2,review3,review4,review5,review6,review7}
The emphasis on quantum critical nature has been summarized
in the famous Doniach's phase diagram,\cite{Doniach}
which has been a guiding principle for a long time
to understand several kinds of anomalous electronic properties
of heavy fermion materials.

The concept of quantum critical point seems to be universal,
since it holds for $p$-, $d$-, and $f$-electron systems.
It is an important issue to accumulate the experimental facts which
can be explained by the universal concept of quantum criticality,
although it is sometimes difficult to control experimentally
quantum criticality in actual materials.
However, if we ignore individual characters of electrons
such as orbital degree of freedom, in general,
difference between $d$ and $f$ electrons cannot be understood
only from the control of the ratio of Coulomb interaction and
electron bandwidth.
It seems to be also important to emphasize individuality of
electron in materials, in particular, when we attempt to synthesize
new exotic and functional materials.

In this context, we are interested in Yb and Tm systems,
which have attracted renewed attention due to difference
in quantum critical nature between Ce and Pr compounds.
For trivalent ions, one and two $f$ electrons are included
in Ce$^{3+}$ and Pr$^{3+}$, respectively,
while one and two $f$ holes exist on Yb$^{3+}$ and Tm$^{3+}$,
respectively.
If we use the electron-hole symmetry,
we expect similar electronic properties
between $f$-electron and hole systems.
Such a discussion may be also found in Pr and Tm compounds.

Thus far, unconventional superconductivity has been found
in Ce-based materials since the pioneering discovery of
superconductivity in CeCu$_2$Si$_2$.\cite{Steglich}
Recently, relatively high superconducting temperature over 2 K has been
also observed in a material group of CeTIn$_5$ (T=Ir, Rh, and Co).
\cite{Ce115-1,Ce115-2,Ce115-3}
Then, if we simply believe the electron-hole picture,
at the first glance,
it seems to be easy to find superconductivity in Yb system.
However, in spite of much effort to seek for superconductivity
in Yb-based heavy-fermion materials,
it has been very difficult to synthesize superconducting Yb compounds.
Recently, superconductivity has been observed in $\beta$-YbAlB$_4$
with a superconducting temperature $T_{\rm c}$=80 mK.
\cite{Nakatsuji1,Nakatsuji2,Nakatsuji3}
It has been claimed that this material exists just on the
quantum critical point at ambient pressure.
In Tm$_5$Rh$_6$Sn$_{18}$, superconductivity has been also
found with $T_{\rm c}$=2.2K.\cite{Kase}
Peculiar reentrant properties have been considered to be
related to the coexistence of magnetism and superconductivity.

If we emphasize similarity between Ce and Yb compounds,
we prefer to exploit the concept of quantum criticality
on the basis of the electron-hole symmetry.
In this case, different quantum critical nature
will be explained by the difference in the local interactions
and the hybridization between conduction and localized $f$ electrons.
However, one may have a simple question
whether quantum criticality of one $f$-hole system is
really the same as that of one $f$-electron system.
In order to clarify $f$-electron state in Yb compounds,
we can choose an alternative way to focus on the difference
in relevant $f$-electron orbital.
This point will be also related to similarity and
difference between Pr- and Tm-based compounds.

For the research along such a direction,
it is necessary to define the $f$-electron state
by a conventional way to include many-body effects.
Here we recall a couple of schemes for the description
of local $f^n$-electron configuration,
where $n$ denotes the number of $f$ electrons
on a localized rare earth or actinide ion.
One is an $LS$ coupling scheme,
in which we construct the spin $\mib{S}$ and angular momentum
$\mib{L}$ by following the Hund's rules as
$\mib{S}$=$\sum_{i=1}^n \mib{s}_i$ and
$\mib{L}$=$\sum_{i=1}^n \mib{\ell}_i$,
where $\mib{s}_i$ and $\mib{\ell}_i$ are spin and angular momenta
for $i$-th $f$ electron, respectively.
As is well known, the Hund's rules are based on the Pauli principle
and Coulomb interactions among $f$ electrons.
After the formation of $\mib{S}$ and ${\mib L}$,
the effect of spin-orbit interaction is included in the form of
$\xi$${\mib L}$$\cdot$${\mib S}$, where $\xi$ is
the spin-orbit coupling in the $LS$ coupling scheme.
We note that $\xi$$>$0 for $n$$<$7, while $\xi$$<$0 for $n$$>$7.
Note also that a good quantum number to label such a state is
the total angular momentum $\mib{J}$,
which is given by $\mib{J}$=$\mib{L}$+$\mib{S}$.
Then, the ground state is characterized by
$J$=$|L$$-$$S|$ for $n$$<$7, while $J$=$L$+$S$ for $n$$>$7.

As is understood from the above discussion,
the $LS$ coupling scheme is quite useful
for the case in which the Hund's rule coupling is much larger
than the spin-orbit interaction,
since ${\mib S}$ and ${\mib L}$ are formed by the Hund's rule coupling
prior to the inclusion of the spin-orbit interaction.
This assumption is considered to be valid
for insulating compounds with localized $f$ electrons.
However, when the spin-orbit interaction is not small compared with
the Hund's rule coupling, for instance in actinide compounds,
the above assumption is not always satisfied.
In addition, if the $f$ electrons begin to be itinerant due to
hybridization with the conduction electrons,
the effect of Coulomb interactions would thereby be effectively reduced.
In rough estimation, the effective size of the Coulomb interaction may
be as large as the bandwidth of $f$ electrons, leading to a violation of
the assumption required for the $LS$ coupling scheme.

For $f$-electron systems in which the spin-orbit interaction becomes
larger than the effective Coulomb interactions,
we prefer to exploit a $j$-$j$ coupling scheme.\cite{Hotta1,Hotta2}
Here we emphasize that the $j$-$j$ coupling scheme is convenient
for the inclusion of many-body effects
by using the standard quantum-field theoretical techniques,
since individual $f$-electron states are clearly defined,
as we explain below.
First, we include the spin-orbit coupling so as to define the state
labelled by the total angular momentum ${\mib j}_i$
for the $i$-th $f$ electron,
given by ${\mib j}_i$=${\mib s}_i$+${\mib \ell}_i$.
For $f$ orbitals with $\ell$=3, we obtain an octet with
$j$=7/2 and a sextet with $j$=5/2,
which are well separated by the spin-orbit interaction.
Note that the level for the octet is higher than that of the sextet.
Then, we consider the effect of Coulomb interactions
to accommodate $n$ electrons among the sextet or octet,
leading to the ground state in the $j$-$j$ coupling scheme.
For the models of Ce and Pr compounds, the sextet should be
used for the construction of the effective model.\cite{Hotta1}
On the other hand, for Yb and Tm materials,
since the sextet is fully occupied,
we consider the octet to construct the model.

In this paper, we develop a prescription to construct
a microscopic effective model for heavy lanthanide systems
such as Yb and Tm compounds on the basis of the $j$-$j$ coupling scheme.
Then, it is shown that the local $f$-electron state in the
$j$-$j$ coupling scheme agrees quite well with that of
the original model including seven $f$ orbitals,
even for a realistic value of the spin-orbit coupling.
Next we consider the impurity Anderson models
to discuss low-temperature multipole states of $f$ electrons.
Here we pick up Yb- and Tm-based filled skutterudites
as typical examples.
The models are analyzed with the use of
a numerical renormalization group technique.
It is found that the multipole state is well reproduced
by the $j$-$j$ coupling scheme
in comparison with that of the seven-orbital Anderson model,
even when we include hybridization between conduction and localized
$f$ electrons for the realistic value of the spin-orbit coupling.

The organization of this paper is as follows.
In Sec.~2, we discuss the local Hamiltonian in the $j$-$j$
coupling scheme in comparison with the results of the
original seven-orbital model.
In Sec.~3, we set the impurity Anderson models
for filled skutterudites.
Then, we show the numerical results for multipole susceptibility
to discuss the validity of the $j$-$j$ coupling model.
In Sec.~4, we provide a few comments on future issues on the present
prescription and summarize this paper.
Throughout this paper, we use such units as $k_{\rm B}$=$\hbar$=1.

%
%
\section{Local Hamiltonian in a $j$-$j$ Coupling Scheme}

\subsection{Original Seven-Orbital Model}

In general, the local $f$-electron Hamiltonian is given by
\begin{equation}
   \label{eq:orgloc}
   H_{\rm loc} = H_{\rm so} + H_{\rm CEF} + H_{\rm int}.
\end{equation}
The first term denotes the spin-orbit coupling, given by
\begin{eqnarray}
   H_{\rm so} = \lambda \sum_{m,\sigma,m',\sigma'}
   \zeta_{m,\sigma;m',\sigma'} f_{m\sigma}^{\dag}f_{m'\sigma'},
\end{eqnarray}
where $\lambda$ is the spin-orbit interaction,
$f_{m\sigma}$ is the annihilation operator of $f$ electron,
$\sigma$=$+1$ ($-1$) for up (down) spin,
$m$ is the $z$-component of angular momentum $\ell$=3,
and the matrix element $\zeta$ is given by
\begin{eqnarray}
  \begin{array}{l}
    \zeta_{m,\sigma;m,\sigma}=m\sigma/2,\\
    \zeta_{m+\sigma,-\sigma;m,\sigma}=\sqrt{\ell(\ell+1)-m(m+\sigma)}/2,
  \end{array}
\end{eqnarray}
and zero for other cases.

The second term denotes crystalline electric field (CEF) potential,
given by
\begin{eqnarray}
  \label{Eq:CEF}
  H_{\rm CEF} = \sum_{m,m',\sigma} B_{m,m'}
  f_{m\sigma}^{\dag} f_{m'\sigma},
\end{eqnarray}
where $B_{m,m'}$ is determined from the CEF table
for $J$=$\ell$=3.\cite{Stevens,Hutchings}
Note that electrostatic CEF potentials do not
act on $f$-electron spin.
Since we will consider later the multipole state of
heavy lanthanide filled skutterudites,
here we show $B_{m,m'}$ of the cubic system with
$T_{\rm h}$ symmetry.\cite{Takegahara}
The results are given by
\begin{eqnarray}
  \begin{array}{l}
    B_{3,3}=B_{-3,-3}=180B_4^0+180B_6^0, \\
    B_{2,2}=B_{-2,-2}=-420B_4^0-1080B_6^0, \\
    B_{1,1}=B_{-1,-1}=60B_4^0+2700B_6^0, \\
    B_{0,0}=360B_4^0-3600B_6^0, \\
    B_{3,-1}=B_{-3,1}=12\sqrt{15}(B_4^4+5B_6^4),\\
    B_{2,-2}=60B_4^4-360B_6^4,\\
    B_{3,1}=B_{-3,-1}=24\sqrt{15}B_6^2,\\
    B_{2,0}=B_{-2,0}=-48\sqrt{30}B_6^2,\\
    B_{1,-1}=360B_6^2,\\
    B_{3,-3}=360B_6^6,\\
  \end{array}
\end{eqnarray}
Note the relation of $B_{m,m'}$=$B_{m',m}$.
We also note the relations of $B_4^4$=$5B_4^0$,
$B_6^4$=$-21B_6^0$, and $B_6^6$=$-B_6^2$.
Following the traditional notation,\cite{LLW}
we define
\begin{eqnarray}
  \begin{array}{l}
    B_4^0=Wx/F(4),\\
    B_6^0=W(1-|x|)/F(6),\\
    B_6^2=Wy/F^t(6),
  \end{array}
\end{eqnarray}
where $x$, $y$, and the sign of $W$ specify the CEF scheme
for $T_{\rm h}$ point group,\cite{Takegahara}
while the absolute value of $W$ determines the energy scale
of the CEF potential.
Concerning non-dimensional parameters, $F(4)$ and $F(6)$,
we choose $F(4)$=15, $F(6)$=180, and $F^t(6)$=24 for $J$=3.

Finally, $H_{\rm int}$ denotes Coulomb interaction term, given by
\begin{equation}
  H_{\rm int} \!=\! \sum_{m_1\sim m_4} \! \sum_{\sigma,\sigma'}
  I_{m_1m_2,m_3m_4}
  f_{m_1\sigma}^{\dag}f_{m_2\sigma'}^{\dag}
  f_{m_3\sigma'}f_{m_4\sigma},
\end{equation}
where the Coulomb integral $I_{m_1m_2,m_3m_4}$ is expressed by
\begin{eqnarray}
  I_{m_1m_2,m_3m_4} = \sum_{k=0}^{6} F^k c_k(m_1,m_4)c_k(m_2,m_3).
\end{eqnarray}
Here $F^k$ is the Slater-Condon parameter \cite{Slater1,Condon}
and $c_k$ is the Gaunt coefficient \cite{Gaunt,Racah}
which is tabulated in the standard textbooks of quantum
mechanics.\cite{Slater2}
Note that the sum is limited by the Wigner-Eckart theorem to
$k$=0, 2, 4, and 6.

\subsection{Effective Hamiltonian for $j$=7/2 Octet}

In order to obtain the model in the $j$-$j$ coupling scheme,
we transform the $f$-electron basis
between $(m,\sigma)$ and $(j,\mu)$ representations,
connected by Clebsch-Gordan coefficients,
where $j$ is the total angular momentum
and $\mu$ is the $z$-component of $j$.
When we define $f_{j\mu}$ as the annihilation operator
for $f$ electron labelled by $j$ and $\mu$,
the transformation is given by
\begin{eqnarray}
  f_{j\mu} = \sum_{m,\sigma} C_{j,\mu;m,\sigma} f_{m\sigma},
\end{eqnarray}
where the Clebsch-Gordan coefficient $C_{j,\mu;m,\sigma}$ is give by
\begin{eqnarray}
 \begin{array}{l}
  C_{5/2,\mu;\mu-\sigma/2,\sigma}=-\sigma \sqrt{(7/2-\sigma \mu)/7},\\
  C_{7/2,\mu;\mu-\sigma/2,\sigma}=\sqrt{(7/2+\sigma \mu)/7},
 \end{array}
\end{eqnarray}
and other components are zero.

After the transformation,
the spin-orbit coupling term is diagonalized as
\begin{equation}
  {\tilde H}_{\rm so} = \sum_{j,\mu}
  {\tilde \lambda}_{j}f^{\dag}_{j\mu}f_{j\mu},
\end{equation}
with ${\tilde \lambda}_{5/2}$=$-2\lambda$
and ${\tilde \lambda}_{7/2}$=$(3/2)\lambda$.
The CEF and Coulomb interaction terms are, respectively,
given by
\begin{equation}
  {\tilde H}_{\rm CEF} = \sum_{j_1\mu_1,j_2\mu_2}
  {\tilde B}^{j_1,j_2}_{\mu_1,\mu_2}
  f^{\dag}_{j_1\mu_1}f_{j_2\mu_2},
\end{equation}
and
\begin{equation}
  {\tilde H}_{\rm int}=\sum_{j_1 \sim j_4} \sum_{\mu_1 \sim \mu_4}
  {\tilde I}^{j_1,j_2; j_3,j_4}_{\mu_1,\mu_2; \mu_3,\mu_4}
  f^{\dag}_{j_1\mu_1}f^{\dag}_{j_2\mu_2}f_{j_3\mu_3}f_{j_4\mu_4},
\end{equation}
where ${\tilde B}$ and ${\tilde I}$ are the CEF potential
and Coulomb interaction, respectively,
in the basis of $j$ and $\mu$.

In the present paper, we consider the model for
heavy lanthanide systems with $n$$>$7.
In the limit of large $\lambda$ for the $j$-$j$ coupling scheme,
$j$=5/2 sextet is fully occupied,
while $j$=7/2 octet is partially occupied.
Thus, here we simply discard all the $j$=5/2 states and
keep only the $j$=7/2 octet.
Namely, we accommodate $n$$-$6 electrons in the $j$=7/2 octet.
Note that in this approximation, the spin-orbit coupling is
given by the effect of potential energy which does not depend
on the orbitals.
Since such an energy can be included in the
chemical potential shift, we do not consider explicitly
${\tilde H}_{\rm so}$ in the following.
Hereafter, we suppress the subscription $j$ in $f_{j\mu}$,
since we consider only the $j$=7/2 octet.

The local model in the $j$-$j$ coupling scheme is given by
\begin{eqnarray}
   \label{eq:jjloc}
   {\tilde H}_{\rm loc} = {\tilde H}_{\rm CEF} + {\tilde H}_{\rm int},
\end{eqnarray}
where ${\tilde H}_{\rm CEF}$ is the CEF potential in the $j$=7/2 octet,
given as
\begin{equation}
  \label{H:CEF}
  {\tilde H}_{\rm CEF} = \sum_{\mu,\nu}
  {\tilde B}_{\mu,\nu} f_{\mu}^{\dag} f_{\nu}.
\end{equation}
Here $f_{\mu}^{\dag}$ is the creation operator
of $f$ electron in the $\mu$-state and $\mu$ indicates the $z$-component of the
total angular momentum which specifies the state in the $j$=7/2 octet.
The CEF potential in the $j$=7/2 octet is given by \cite{Hutchings}
\begin{eqnarray}
  \label{jj-CEF}
  \begin{array}{l}
    {\tilde B}_{7/2,7/2}={\tilde B}_{-7/2,-7/2}
    =420{\tilde B}_4^0+1260 {\tilde B}_6^0, \\
    {\tilde B}_{5/2,5/2}={\tilde B}_{-5/2,-5/2}
    =-780{\tilde B}_4^0-6300 {\tilde B}_6^0, \\
    {\tilde B}_{3/2,3/2}={\tilde B}_{-3/2,-3/2}
    =-180{\tilde B}_4^0+11340{\tilde B}_6^0, \\
    {\tilde B}_{1/2,1/2}={\tilde B}_{-1/2,-1/2}
    =540{\tilde B}_4^0-6300 {\tilde B}_6^0, \\
    {\tilde B}_{7/2,-1/2}=B_{-7/2,1/2}
    =12\sqrt{35}({\tilde B}_4^4+15{\tilde B}_6^4),\\
    {\tilde B}_{5/2,-3/2}=B_{-5/2,3/2}
    =60\sqrt{3}({\tilde B}_4^4-7{\tilde B}_6^4),\\
    {\tilde B}_{7/2,3/2}=B_{-7/2,-3/2}=120\sqrt{21}{\tilde B}_6^2,\\
    {\tilde B}_{5/2,1/2}=B_{-5/2,-1/2}=-504\sqrt{5}{\tilde B}_6^2,\\
    {\tilde B}_{3/2,-1/2}=B_{-3/2,1/2}=168\sqrt{15}{\tilde B}_6^2,\\
    {\tilde B}_{7/2,-5/2}=B_{-7/2,5/2}=360\sqrt{7}{\tilde B}_6^6,\\
  \end{array}
\end{eqnarray}
where we note again the relations of
${\tilde B}_4^4$=$5{\tilde B}_4^0$,
${\tilde B}_6^4$=$-21{\tilde B}_6^0$,
and ${\tilde B}_6^6$=$-{\tilde B}_6^2$.
The CEF parameters for $j$=7/2 are related to those for $J$=$\ell$=3 as
\begin{eqnarray}
  \begin{array}{l}
    {\tilde B}_4^0=(\beta_{7/2}/\beta_{3})B_4^0=3B_4^0/7,\\
    {\tilde B}_6^0=(\gamma_{7/2}/\gamma_{3})B_6^0=B_6^0/7,\\
    {\tilde B}_6^2=(\gamma_{7/2}/\gamma_{3})B_6^2=B_6^2/7,
  \end{array}
\end{eqnarray}
where $\beta_J$ and $\gamma_J$ are fourth- and sixth-order
Stevens factors, respectively.\cite{Stevens}
We note that $\beta_{3}$=$2/495$, $\beta_{7/2}$=$2/1155$,
$\gamma_{3}$=$-4/3861$, and $\gamma_{7/2}$=$-4/27027$.

The second term in eq.~(\ref{eq:jjloc})
indicates the Coulomb interactions in the $j$=7/2 octet,
which is given by
\begin{equation}
  {\tilde H}_{\rm int}=\sum_{\mu,\nu,\mu',\nu'}
  {\tilde I}_{\mu,\nu;\nu',\mu'}
  f_{\mu}^{\dag} f_{\nu}^{\dag}f_{\nu'} f_{\mu'},
\end{equation}
where ${\tilde I}$ is the matrix element for Coulomb interactions
among $j$=7/2 states.
In order to classify the Coulomb interactions in the $j$=7/2 octet,
we consider the situation where we accommodate two electrons in the octet.
Note that the allowed values for total angular momentum $J$ are 0, 2, 4, and 6
due to the Pauli principle.
Thus, the Coulomb interaction term should be written
in a 28$\times$28 matrix form.
Note that ``28'' is the sum of the basis numbers for singlet ($J$=0),
quintet ($J$=2), nonet ($J$=4), and tridectet ($J$=6).
As is easily understood,
this 28$\times$28 matrix can be decomposed into a block-diagonalized
form labelled by $J_z$, including
one 4$\times$4 matrix for $J_z$=0,
four 3$\times$3 matrices for $J_z$=$\pm$1 and $\pm$2,
four 2$\times$2 matrices for $J_z$=$\pm$3 and $\pm$4,
and four 1$\times$1 for $J_z$=$\pm$5 and $\pm$6.
We skip the details of tedious calculations for the evaluation of
matrix elements and show only the results by using
the parameters $E_k$ ($k$=0,1,2,3),\cite{Flowers1,Flowers2}
which are related to the Slater-Condon parameters $F^k$ as \cite{note}
\begin{equation}
  \begin{array}{l}
    E_0 = F^0-{25 \over 567}F^2-{5 \over 231}F^4-{125 \over 11583}F^6,\\
    E_1 = {40 \over 567}F^2+{8 \over 231}F^4+{200 \over 11583}F^6,\\
    E_2 = {2 \over 1617}F^2-{2 \over 5929}F^4,\\
    E_3 = {10 \over 43659}F^2+{2 \over 17787}F^4-{100 \over 1656369}F^6.
  \end{array}
\end{equation}
For $J_z$=6 and 5, we obtain
\begin{equation}
  \label{Eq:Jz6}
    {\tilde I}_{7/2,5/2;5/2,7/2}=E_0-154E_3,
\end{equation}
and
\begin{equation}
  \label{Eq:Jz5}
    {\tilde I}_{7/2,3/2;3/2,7/2}=E_0-154E_3,
\end{equation}
respectively.
For $J_z$=4 and 3, we obtain
\begin{equation}
  \label{Eq:Jz4}
  \begin{array}{l}
    {\tilde I}_{7/2,1/2;1/2,7/2} = E_0-35E_2/2-119E_3/2, \\
    {\tilde I}_{5/2,3/2;3/2,5/2} = E_0-75E_2/2+97E_3/2, \\
    {\tilde I}_{7/2,1/2;3/2,5/2} = \sqrt{105}(5E_2-27E_3)/2,
  \end{array}
\end{equation}
and
\begin{equation}
  \label{Eq:Jz3}
  \begin{array}{l}
    {\tilde I}_{7/2,-1/2;-1/2,7/2} = E_0-35E_2+35E_3, \\
    {\tilde I}_{5/2, 1/2; 1/2,5/2} = E_0-20E_2-46E_3, \\
    {\tilde I}_{7/2,-1/2; 1/2,5/2} = \sqrt{7}(10E_2-54E_3).
  \end{array}
\end{equation}
For $J_z$=2 and 1, we obtain
\begin{equation}
  \label{Eq:Jz2}
  \begin{array}{l}
    {\tilde I}_{7/2,-3/2;-3/2,7/2} = E_0-21E_2+98E_3, \\
    {\tilde I}_{5/2,-1/2;-1/2,5/2} = E_0+35E_2-46E_3, \\
    {\tilde I}_{3/2, 1/2; 1/2,3/2} = E_0+30E_2+80E_3, \\
    {\tilde I}_{7/2,-3/2;-1/2,5/2} = -\sqrt{105}(2E_2+9E_3),\\
    {\tilde I}_{7/2,-3/2; 1/2,3/2} = 9\sqrt{35}(E_2-E_3),\\
    {\tilde I}_{5/2,-1/2; 1/2,3/2} = -\sqrt{3}(25E_2+63E_3),
  \end{array}
\end{equation}
and
\begin{equation}
  \label{Eq:Jz1}
  \begin{array}{l}
    {\tilde I}_{7/2,-5/2;-5/2,7/2} = E_0+49E_2/2+259E_3/2, \\
    {\tilde I}_{5/2,-3/2;-3/2,5/2} = E_0+27E_2+17E_3, \\
    {\tilde I}_{3/2,-1/2;-1/2,3/2} = E_0-15E_2/2-29E_3/2, \\
    {\tilde I}_{7/2,-5/2;-3/2,5/2} = -\sqrt{21}(13E_2+9E_3),\\
    {\tilde I}_{7/2,-5/2;-1/2,3/2} = 9\sqrt{105}(E_2-E_3)/2,\\
    {\tilde I}_{5/2,-3/2;-1/2,3/2} = -3\sqrt{5}(E_2+21E_3),
  \end{array}
\end{equation}
Finally, for $J_z$=0, we obtain
\begin{equation}
  \label{Eq:Jz0}
  \begin{array}{l}
    {\tilde I}_{7/2,-7/2;-7/2,7/2} = E_0+E_1+49E_2+105E_3, \\
    {\tilde I}_{3/2,-3/2;-3/2,3/2} = E_0+E_1-29E_2+ 51E_3, \\
    {\tilde I}_{5/2,-5/2;-5/2,5/2} = E_0+E_1+ 9E_2- 75E_3, \\
    {\tilde I}_{1/2,-1/2;-1/2,1/2} = E_0+E_1+15E_2+ 51E_3, \\
    {\tilde I}_{7/2,-7/2;-5/2,5/2} =-E_1- 49E_2/2+ 49E_3/2, \\
    {\tilde I}_{7/2,-7/2;-3/2,3/2} = E_1- 21E_2  - 56E_3, \\
    {\tilde I}_{7/2,-7/2;-1/2,1/2} =-E_1+105E_2/2+ 49E_3/2, \\
    {\tilde I}_{5/2,-5/2;-3/2,3/2} =-E_1+ 21E_2/2-131E_3/2, \\
    {\tilde I}_{5/2,-5/2;-1/2,1/2} = E_1+ 15E_2/2- 92E_3, \\
    {\tilde I}_{3/2,-3/2;-1/2,1/2} =-E_1- 45E_2/2-131E_3/2, \\
  \end{array}
\end{equation}
Note here the following relations:
\begin{equation}
  {\tilde I}_{\mu,\nu;\nu',\mu'}={\tilde I}_{\mu',\nu';\nu,\mu},
\end{equation}
and
\begin{equation}
  {\tilde I}_{\mu,\nu;\nu',\mu'}={\tilde I}_{-\nu,-\mu;-\mu',-\nu'}.
\end{equation}
By using these two relations and eqs.~(\ref{Eq:Jz6})-(\ref{Eq:Jz0}), 
we can obtain all the Coulomb matrix elements.

\subsection{CEF Energy Levels}

Let us now consider the situation in which two electrons are
accommodated in the $j$=7/2 octet.
This situation indicates the case with 8 electrons in $f$ orbitals,
corresponding to $f^8$ configuration of Tb$^{3+}$ ion.
When we diagonalize the 28$\times$28 matrix
for Coulomb interaction terms,
we can easily obtain the eigen energies as
$E_0-154E_3$ for the $J$=6 tridectet,
$E_0-55E_2+143E_3$ for the $J$=4 nonet,
$E_0+99E_2+143E_3$ for the $J$=2 quintet,
and $E_0+4E_1$ for the $J$=0 singlet.
These values are exactly the same as those obtained
in the nuclear shell theory in the $j$-$j$ coupling scheme.
\cite{Talmi}
For typical values of Slater-Condon parameters, we find that
the ground state is specified by $J$=6 in the $j$-$j$ coupling scheme.
For Tb$^{3+}$ ion, in the $LS$ coupling scheme,
we obtain the ground-state level as $^{7}F$ with $S$=3 and $L$=3
from the Hund's rules.
On further inclusion of the spin-orbit interaction,
the ground state becomes characterized by $J$=6,
expressed as $^7F_6$ in the traditional notation.
Note that we are considering a two-electron problem.
Thus, when we correctly include the effects of Coulomb interactions,
the same quantum number as that in the $LS$ coupling scheme
is obtained in the $j$-$j$ coupling scheme
for the ground-state multiplet.

In order to discuss the CEF energy levels, it is necessary to
determine the values of local interactions.
Among them, concerning the Slater-Condon parameters,
we set $F^0$=10 eV by hand.
The magnitude of $F^0$ is related to the absolute value of
the ground state energy.
It can be evaluated by the first-principles calculation,
but it is out of the scope of the present paper.
Other Slater-Condon parameters are determined so as
to reproduce excitation spectra of Pr$^{3+}$ ion.\cite{Cai,Eliav}
After the fitting, we obtain
$F^2$=8.75 eV, $F^4$=6.60 eV, and $F^6$=4.44 eV.\cite{Hotta3,Hotta4}
As long as we ignore the difference in lanthanide ions,
e.g., the size of ion radius,
we use these values for all lanthanide ions.
On the other hand, as for the spin-orbit coupling $\lambda$,
we use the value which has been determined experimentally
for each lanthanide ion.\cite{spin-orbit}

\begin{figure}[t]
\begin{center}
\includegraphics[width=8.5truecm]{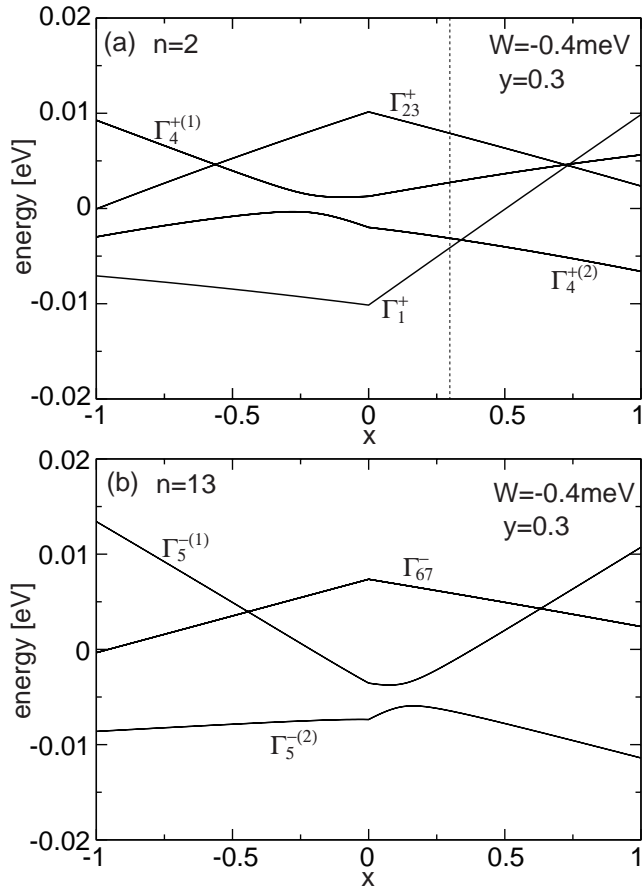}
\caption{CEF energy levels for (a) $n$=2 and (b) $n$=13.
The vertical line denote the position of $x$=0.3.
Concerning other parameters, see the main text.
}
\end{center}
\end{figure}

Concerning CEF parameters, it is necessary to specify the
actual material, since they depend on the crystal structure
and the kinds of ligand ions.
Here we consider the case of filled skutterudite materials,
since we will analyze the multipole state of filled skutterudites
later in this paper.
The CEF parameters are different from material to material
even if we fix the material group,
but the typical values are $W$=$-$0.4 meV, $y$=0.3, and $x$=0.3,
which are determined
so as to reproduce quasi-quartet CEF scheme of PrOs$_4$Sb$_{12}$.
\cite{Kohgi,Kuwahara,Goremychkin}
Note that for Pr atom, we use $\lambda$=0.095 eV
from the experimental value.

In Fig.~1(a), we show the results of CEF energy levels vs. $x$
for the case of $n$=2, corresponding to Pr$^{3+}$,
with the use of the above parameters.
The vertical dash line denotes the position of $x$=0.3
and we can understand that the ground state is $\Gamma_1^+$ singlet
and the first excited state is $\Gamma_4^{+(2)}$ triplet
with the small excitation energy.
This is considered to be a typical situation of PrOs$_4$Sb$_{12}$.
When we change the values of $x$ and/or $y$,
we can obtain another situation
for different filled skutterudite material.

In Fig.~1(b), we show the CEF energy levels vs. $x$
for the case of $n$=13 with $W$=$-$0.4 meV and $y$=0.3.
The spin-orbit coupling $\lambda$ is set as 0.356 eV
for Yb atom.\cite{spin-orbit}
In the $O_{\rm h}$ point group,
the octet is known to split into two doublets and one quartet.
\cite{LLW}
In the $T_{\rm h}$ point group, on the other hand,
two doublets are mixed.\cite{Takegahara}
In the present parameters, we always find the doublet ground state,
irrespective of the values of $x$.

\begin{figure}[t]
\begin{center}
\includegraphics[width=8.5truecm]{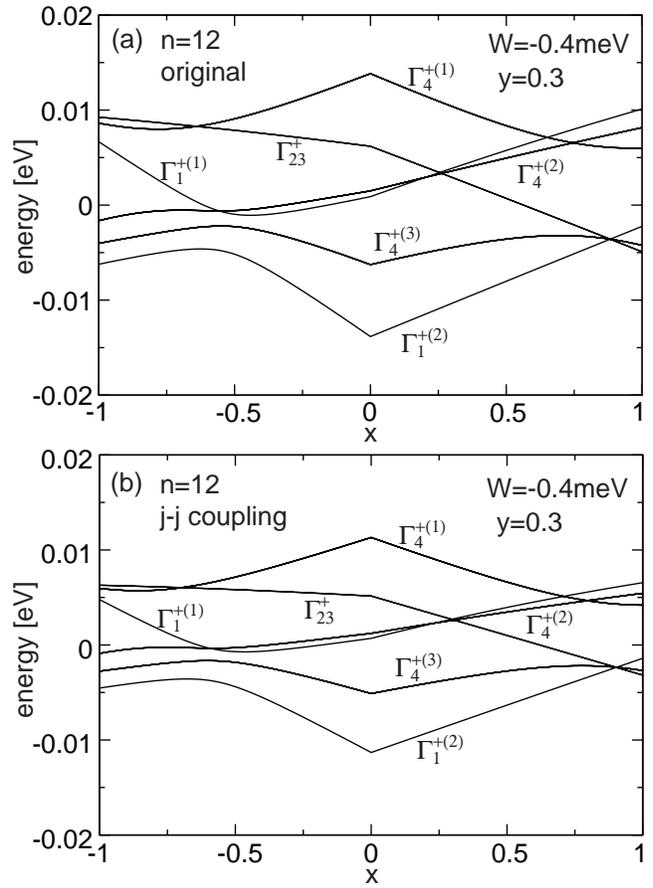}
\caption{CEF energy levels vs. $x$ for $n$=12
(a) in the original seven-orbital model eq.~(\ref{eq:orgloc})
and (b) in the $j$-$j$ coupling model eq.~(\ref{eq:jjloc}).
}
\end{center}
\end{figure}

In Figs.~2, we show the results for the case of $n$=12
corresponding to Tm$^{3+}$ ion,
in order to see the validity of the $j$-$j$ coupling scheme.
In Fig.~2(a), the result for the original seven-orbital model
in eq.~(\ref{eq:orgloc}) is shown.
In the $O_{\rm h}$ group,
the tridectet of $J$=6 is split into two singlets, one doublet, and
three triplets.\cite{LLW}
In the $T_{\rm h}$ group, on the other hand, two singlets are mixed.
Three triplets are also mixed due to the $y$-term.
In the present parameters, the ground state around at $x$=0.3
is characterized by $\Gamma_1^{+}$ singlet.
However, when we increase the value of $x$, we find the change of
the ground state from $\Gamma_1^+$ singlet to $\Gamma_{23}^{+}$
non-Kramers doublets.
Such a change is found to occur around at $x \sim 0.9$.

In Fig.~2(b), we show the CEF energies vs. $x$ of the
$j$-$j$ coupling model eq.~(\ref{eq:jjloc})
with the use of the same parameters,
except for the value of the spin-orbit coupling.
In the $j$-$j$ coupling scheme, $\lambda$ is set as infinity.
Nevertheless, in the first impression,
even if the actual value of $\lambda$ is finite,
the results of the $j$-$j$ coupling model agree quite well
with those of the original seven-orbital model in Fig.~2(a).
If we effectively change the absolute value of $W$ as a
fitting parameter, Fig.~2(a) can be reproduced by the
$j$-$j$ coupling model quantitatively.

The reason why the CEF energy levels are well reproduced by
the $j$-$j$ coupling scheme even for the finite value of $\lambda$
is as follows.
Since the maximum value of the difference in the $z$-component of
total angular momentum is seven among $j$=7/2 states,
the sixth-order CEF potential can be included
in the $j$-$j$ coupling scheme,
in sharp contrast to the case of $j$=5/2.\cite{Hotta1}
Thus, the difference in the value of $\lambda$ does not provide
serious effect on the CEF ground state,
as long as we consider $\lambda$$\gg$$|W|$.
In fact, after lengthy algebraic calculations, we obtain
the overlap integral of the ground states for $n$=12
as $\langle \Phi | \Phi_{j-j} \rangle$=$\sqrt{6/7}$=$0.926$,
\cite{Flowers1,Flowers2}
where $|\Phi \rangle$ and  $|\Phi_{j-j} \rangle$ denote
the CEF ground state of the seven-orbital model and
the $j$-$j$ coupling one, respectively.

We do not show further the results for the cases of $n$$<$12,
but the CEF states of the $j$-$j$ coupling model can reproduce
well those of the original seven-orbital model.
Thus, we conclude that the local $f$-electron state of
heavy lanthanide systems is well approximated
by the $j$-$j$ coupling scheme.
This is one of important messages of the present paper.

%
%
\section{Numerical Analysis of Impurity Anderson Models}

We have explained the prescription to obtain the local Hamiltonian
in the $j$-$j$ coupling scheme.
Even if we use only the $j$=7/2 octet, it is possible to reproduce
local multi-$f$-electron state, which agrees well with those
obtained in the original seven-orbital model.
By including further the itinerancy of $f$ electrons,
we can discuss magnetism and superconductivity
of heavy lanthanide compounds from a microscopic viewpoint.
In this section, as an example, we discuss the multipole state
of Yb- and Tm-based filled skutterudites
by using the Anderson models.
Then, we show the effectiveness of the $j$-$j$ coupling model
in the microscopic level.

\subsection{Anderson Models}

The seven-orbital Anderson Hamiltonian is given by
\begin{equation}
  \label{And-org}
    H \!=\! \sum_{\mib{k},\sigma}
    \varepsilon_{\mib{k}} c_{\mib{k}\sigma}^{\dag} c_{\mib{k}\sigma}
    \!+\! \sum_{\mib{k},\sigma,m}
    (V_{m} c_{\mib{k}\sigma}^{\dag}f_{m\sigma}+{\rm h.c.}) \!+\! H_{\rm loc},
\end{equation}
where $\varepsilon_{\mib{k}}$ denotes conduction electron dispersion,
$c_{\mib{k}\sigma}$ indicates the annihilation operator for conduction
electron with momentum $\mib{k}$ and spin $\sigma$,
$V_{m}$ is the hybridization between conduction and $f$ electrons,
and the local $f$-electron term $H_{\rm loc}$ is already
given in eq.~(\ref{eq:orgloc}).
For filled skutterudites, the main conduction band is given by
$a_{\rm u}$, constructed from $p$-orbitals of pnictogen.\cite{Harima}
Note that the hybridization occurs between the states
with the same symmetry.
Since the $a_{\rm u}$ conduction band has xyz symmetry,
we set $V_2$=$V/\sqrt{2}$, $V_{-2}$=$-V/\sqrt{2}$,
and zeros for other values of $m$.
The hybridization is fixed as $V$=0.05 eV and a half of the bandwidth
of $a_{\rm u}$ conduction band is set as 1 eV.

The $j$-$j$ coupling Anderson model is given by
\begin{equation}
    \label{And-jj}
    {\tilde H} \!=\! \sum_{\mib{k},\sigma}
    \varepsilon_{\mib{k}} c_{\mib{k}\sigma}^{\dag} c_{\mib{k}\sigma}
    \!+\! \sum_{\mib{k},\sigma,\mu}
    ({\tilde V}_{\sigma,\mu} c_{\mib{k}\sigma}^{\dag}f_{\mu}+{\rm h.c.})
    \!+\! {\tilde H}_{\rm loc},
\end{equation}
where the local $f$-electron term in the $j$-$j$ coupling scheme
${\tilde H}_{\rm loc}$ is given by eq.~(\ref{eq:jjloc}).
Since the $a_{\rm u}$ conduction band has xyz symmetry,
which is described by $\Gamma_5^{-}$ in the $T_{\rm h}$ group,
i.e., $\Gamma_7^{-}$ in the $O_{\rm h}$ group,
we set ${\tilde V}_{\uparrow,5/2}$=$V\sqrt{3}/2$,
${\tilde V}_{\uparrow,-3/2}$=$-V/2$,
${\tilde V}_{\downarrow,-5/2}$=$-V\sqrt{3}/2$,
${\tilde V}_{\downarrow,3/2}$=$V/2$,
and zeros for other cases.
Note that the connectivity between $\mu$ and $\sigma$
is determined by the definition of pseudo-spin of $f$-electron state
on the basis of the time reversal symmetry.
For the $j$-$j$ coupling model, we also set $V$=0.05 eV for simplicity,
since this value is not so important at low enough temperatures
in the following discussion.
Note that for the $T_{\rm h}$ group for filled skutterudites,
$\Gamma_6^{-}$ and $\Gamma_7^{-}$ doublets in the $O_{\rm h}$ group
are mixed and they are expressed as $\Gamma_5^{-(1)}$ and $\Gamma_5^{-(2)}$.

It should be noted that
we do not show explicitly the chemical potential terms both
in the models, but in actual calculations, we set the value of the
chemical potential so as to fix the local $f$-electron number
as $n$=13 or 12.

\subsection{Multipole Operator}

In order to discuss the multipole state,
it is necessary to define the multipole operator.
The details can be found in Refs.~\citen{Hotta3} and \citen{Hotta4},
but here we briefly explain the method to define the multipole operator
to make this paper self-contained with some additional comments.

When we consider multipole operator for $f$ electrons,
it should be defined in the one-body form
as an extension of charge and total angular momentum operators
on the basis of a belief that the multipole denotes
the combined degree of freedom of spin and orbital.
\cite{Hotta4}

In the multipole expansion of potential in electromagnetism,
higher electric and magnetic multipole moments appear
in the coefficients of the expansion by
the spherical harmonics $Y_{LM}$ with larger angular momentum.
In group theory, $Y_{LM}$ is defined by the basis of
irreducible representation $D^{(L)}$ of the rotation group $R$,
expressed as
\begin{equation}
  R Y_{LM}=\sum_{M'} Y_{LM'} D_{MM'}^{(L)}.
\end{equation}
In order to define $f$-electron multipole operator,
on the analogy of the multipole expansion,
we exploit a concept of spherical tensor operator
in the quantum mechanics of angular momentum.\cite{Inui}
When we consider the rotation of operator ${\hat T}$,
we obtain a set of operators ${\hat T}^{(k)}$=$\{ {\hat T}_q^{(k)} \}$
with $(2k+1)$-components ($q=-k, -k+1, \cdots, k-1, k$), given by
\begin{equation}
  R {\hat T}_{q}^{(k)} R^{-1}
  =\sum_{q'} {\hat T}_{q'}^{(k)}D_{qq'}^{(k)}.
\end{equation} 
Namely, ${\hat T}_q^{(k)}$ is transformed like a basis of
irreducible representation $D^{(k)}$ for the rotation.
Such ${\hat T}_q^{(k)}$ is called spherical tensor operator of rank $k$.

Thus far, we have implicitly assumed $f$-electron density
in an isolated ion, but in actuality, rare-earth ions are
put in the crystal structure.
Then, it is convenient to change from spherical to cubic
tensor operators, given by
\begin{equation}
  {\hat T}^{(k)}_{\gamma}
  =\sum_q G^{(k)}_{\gamma,q}{\hat T}^{(k)}_{q},
\end{equation}
where $k$ is a rank of multipole,
an integer $q$ runs between $-k$ and $k$,
$\gamma$ is a label to express $O_{\rm h}$ irreducible representation,
and $G^{(k)}_{\gamma,q}$ is the transformation matrix
between spherical and cubic harmonics.
Then, the cubic tensor operator 
for $f$ electron is expressed in the second-quantized form as
\begin{equation}
  {\hat T}^{(k)}_{\gamma} = \sum_{m\sigma,m'\sigma'}
  T^{(k,\gamma)}_{m\sigma,m'\sigma'}f^{\dag}_{m\sigma}f_{m'\sigma'}.
\end{equation}
Throughout this paper, we use the cubic tensor operator as multipole.

As for the classification of multipole,
we use the notations in the group theory.
We express the irreducible representation of the CEF state
by Bethe notation in this paper, but for multipoles,
we use short-hand notations by the combination of the number
of irreducible representation and the parity of time reversal
symmetry, g for gerade and u for ungerade.
Note also that for the $T_{\rm h}$ group,
$\Gamma_1$ and $\Gamma_2$ of $O_{\rm h}$ are mixed.
We remark that $\Gamma_4$ and $\Gamma_5$ of $O_{\rm h}$ are
also mixed in $T_{\rm h}$.
Thus, we obtain six independent multipole components as
1g+2g, 2u, 3g, 3u, 4g+5g, and 4u+5u for filled skutterudites.
Note that 1u does not appear within rank 7.

The coefficient $T^{(k,\gamma)}_{m\sigma,m'\sigma'}$ is
calculated from the spherical tensor operator as follows.
First we change the $f$-electron basis
from $(m,\sigma)$ to $(j,\mu)$.
Note that $j$ takes $7/2$ and $5/2$ for $f$ electrons.
For a certain value of angular momentum $j$ and its $z$-component $\mu$,
the matrix element of spherical tensor operator is
easily calculated by the Wigner-Eckart theorem as
\begin{equation}
  \langle j \mu | T_q^{(k)} | j\mu' \rangle
  =\frac{\langle j || T^{(k)} || j \rangle}{\sqrt{2j+1}}
  \langle j\mu | j\mu' kq \rangle,
\end{equation}
where $\langle JM | J'M' J'' M'' \rangle$ denotes
the Clebsch-Gordan coefficient and
$\langle j || T^{(k)} || j \rangle$ is
the reduced matrix element for spherical tensor operator,
given by
\begin{equation}
  \label{red}
  \langle j || T^{(k)} || j \rangle=
  \frac{1}{2^k} \sqrt{\frac{(2j+k+1)!}{(2j-k)!}}.
\end{equation}
Note that $k \le 2j$ and the highest rank is $2j$.
The coefficient $T^{(k,q)}_{m\sigma,m'\sigma'}$ is obtained
by returning to the basis of $(m,\sigma)$ from $(j,\mu)$.
The final result is given by
\begin{equation}
  \label{Tkq}
  \begin{split}
    T^{(k,\gamma)}_{m\sigma,m'\sigma'}
    &= \sum_{j,\mu,\mu',q}G^{(k)}_{\gamma,q}
    \frac{\langle j || T^{(k)} || j \rangle}{\sqrt{2j+1}}
    \langle j \mu | j \mu' k q \rangle \\
    &\times
    \langle j \mu | \ell m s \frac{\sigma}{2} \rangle
    \langle j \mu' | \ell m' s \frac{\sigma'}{2} \rangle,
  \end{split}
\end{equation}
where $\ell$=3, $s$=1/2, $j$=$\ell$$\pm$$s$,
and $\mu$ runs between $-j$ and $j$.

For the $j$-$j$ coupling scheme in the $j$=7/2 octet,
we should discard the contribution from $j$=5/2 sextet.
Then, the multipole operator in the $j$-$j$ coupling scheme
is expressed in the second-quantized form as
\begin{equation}
  {\hat T}^{(k)}_{\gamma} = \sum_{\mu,\mu'}
  {\tilde T}^{(k,\gamma)}_{\mu,\mu'}f^{\dag}_{\mu}f_{\mu'},
\end{equation}
where $\mu$ denotes the $z$-component of $j$=7/2.
The coefficient ${\tilde T}^{(k,q)}_{\mu,\mu'}$ is
given by
\begin{equation}
  {\tilde T}^{(k,\gamma)}_{\mu,\mu'}
  =\sum_{q} G^{(k)}_{\gamma,q}
  \frac{\langle j || T^{(k)} || j \rangle}{\sqrt{2j+1}}
  \langle j \mu | j \mu' k q \rangle,
\end{equation}
where $j$ is fixed as $j$=7/2 in this equation.
We use this definition for the calculation of
the multipole susceptibility in the $j$-$j$ coupling scheme.

It should be noted here that multipoles belonging to the same symmetry
are mixed in general, even if the rank is different.
In addition, multipoles are also mixed due to the effect of
CEF potentials of the $T_{\rm h}$ group.
Namely, the $f$-electron spin-charge density should be given by
the appropriate superposition of multipoles,
expressed as
\begin{equation}
  \label{multi}
  {\hat X}=\sum_{k,\gamma} p^{(k)}_{\gamma}{\hat T}^{(k)}_{\gamma}.
\end{equation} 
In order to determine the coefficient $p^{(k)}_{\gamma}$,
it is necessary to evaluate the multipole susceptibility
in the linear response theory.
However, multipoles belonging to the same symmetry are mixed in general,
even if the rank is different.
In addition, multipoles are also mixed due to the CEF effect.
Thus, it is natural to define $p^{(k)}_{\gamma}$ by the eigenstate
of susceptibility matrix
\begin{equation}
  \label{sus}
  \begin{split}
    \chi_{k\gamma,k'\gamma'} \!
    = & \frac{1}{Z}
    \sum_{i,j} \frac{e^{-E_i/T}-e^{-E_j/T}}{E_j-E_i}
    \langle i | [{\hat T}^{(k)}_{\gamma}-\rho^{(k)}_{\gamma}] | j \rangle
    \\ & \times 
    \langle j | [{\hat T}^{(k')}_{\gamma'}- \rho^{(k')}_{\gamma'}]| i \rangle,
  \end{split}
\end{equation}
where $E_i$ is the eigenenergy for the $i$-th eigenstate
$|i\rangle$ of $H$ or ${\tilde H}$, $T$ is a temperature,
$\rho^{(k)}_{\gamma}$=$\sum_i e^{-E_i/T}
\langle i |{\hat T}^{(k)}_{\gamma}| i \rangle/Z$,
and $Z$ is the partition function given by
$Z$=$\sum_i e^{-E_i/T}$.
Note that the multipole susceptibility is given by
the eigenvalue of the susceptibility matrix.

\subsection{Method}

In order to evaluate the multipole susceptibility of the impurity
Anderson model, here we employ
a numerical renormalization group (NRG) method.\cite{NRG}
In this technique, we can include efficiently the conduction
electrons states near the Fermi energy by discretizing
momentum space logarithmically.
Note that in actual calculations, it is necessary
to introduce a cut-off $\Lambda$ for
the logarithmic discretization of the conduction band.
Due to the limitation of computer resources,
we keep only $M$ low-energy states.
In this paper, we set $\Lambda$=6 and $M$=2000.
The temperature $T$ is defined as
$T$=$\Lambda^{-(N-1)/2}$ in the NRG calculation,
where $N$ is the number of the renormalization step.
With the use of NRG technique, we evaluate
entropy $S_{\rm imp}$, specific heat $C_{\rm imp}$,
and multipole susceptibility $\chi$.
In particular, the optimized multipole state
is defined by the eigen state with the maximum eigen value
of multipole susceptibility matrix eq.~(\ref{sus}).

\subsection{Results}

\begin{figure}[t]
\begin{center}
\includegraphics[width=8.5truecm]{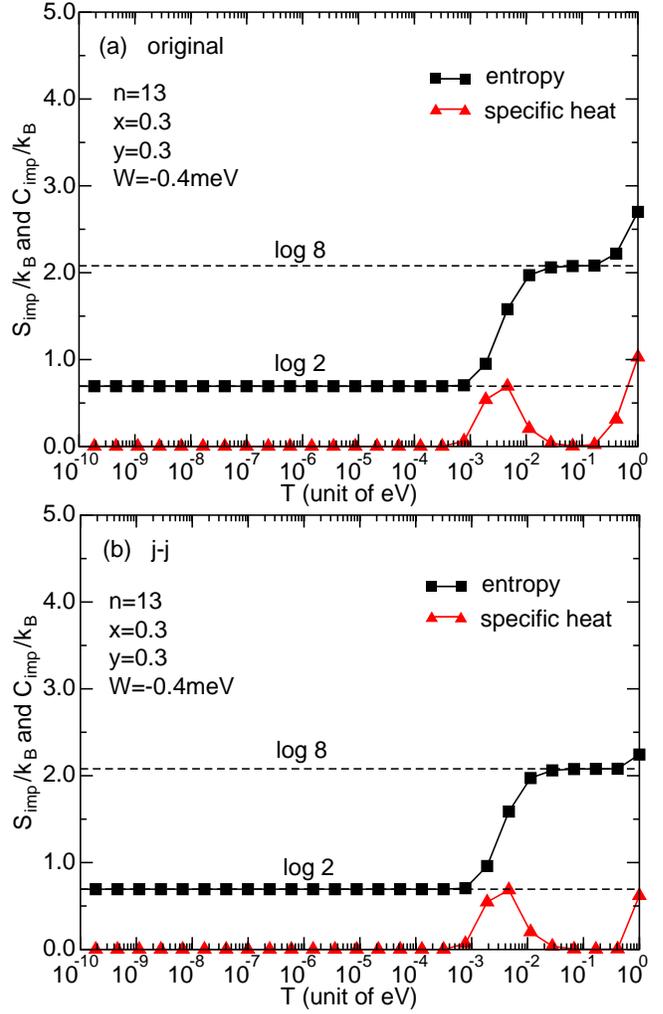}
\caption{(Color online) Entropy $S_{\rm imp}$ and specific heat $C_{\rm imp}$ of
(a) the original seven-orbital model eq.~(\ref{And-org}) and
(b) the $j$-$j$ coupling model eq.~(\ref{And-jj})
for $n$=13 and $x$=0.3.
}
\end{center}
\end{figure}

Let us now show our numerical results.
First we consider the case of $n$=13.
In Figs.~3, we depict the results of entropy and specific heat
both for the original seven-orbital Anderson model eq.~(\ref{And-org})
and the $j$-$j$ coupling Anderson model eq.~(\ref{And-jj}).
Except for the high-temperature region such as $T$$\sim$1 eV,
we do not find significant difference between two panels.
Since the local $f$-electron number is fluctuating
due to the effect of hybridization, the $f$-electron wave function of
the original seven-orbital Anderson model is not equal to
that of the $j$-$j$ coupling Anderson model.
However, as naively expected from the similarity in the
local $f$-electron states, the $j$-$j$ coupling model
can reproduce well the results of the original model.

The difference between both models can be observed at high temperatures
as $T$$\sim$1 in the region without enough renormalization steps.
Since in the $j$-$j$ coupling model, we discard the $j$=5/2 states,
it is natural that there appears difference from the original
seven-orbital model at high temperatures.
Here the energy unit is a half of the conduction bandwidth,
which is in the order of eV.
Since $\lambda$ is in the order of 0.1 eV, it is reasonable
that the deviation can be found even at $T$$>$0.1.

Around at a temperature of 0.1, we find a plateau in the entropy
with the value of $\log 8$.
It is easily understood that this is due to 8-fold degeneracy
of $j$=7/2, since seven electrons (or one hole) are included
in the $j$=7/2 octet.
A part of the entropy $\log 8$ is released around
at $T$=0.01$\sim$0.001 and
a peak in the specific heat is found at $T$=0.005.
Then, we find residual entropy of $\log 2$ in the temperature
region less than $T$=0.001.
The appearance of the partial entropy release is understood
due to the CEF energy splitting.
Namely, as observed in Fig.~1(b), the octet is found to be
split in the CEF energy scale in the order of 0.01 eV.

\begin{figure}[t]
\begin{center}
\includegraphics[width=8.5truecm]{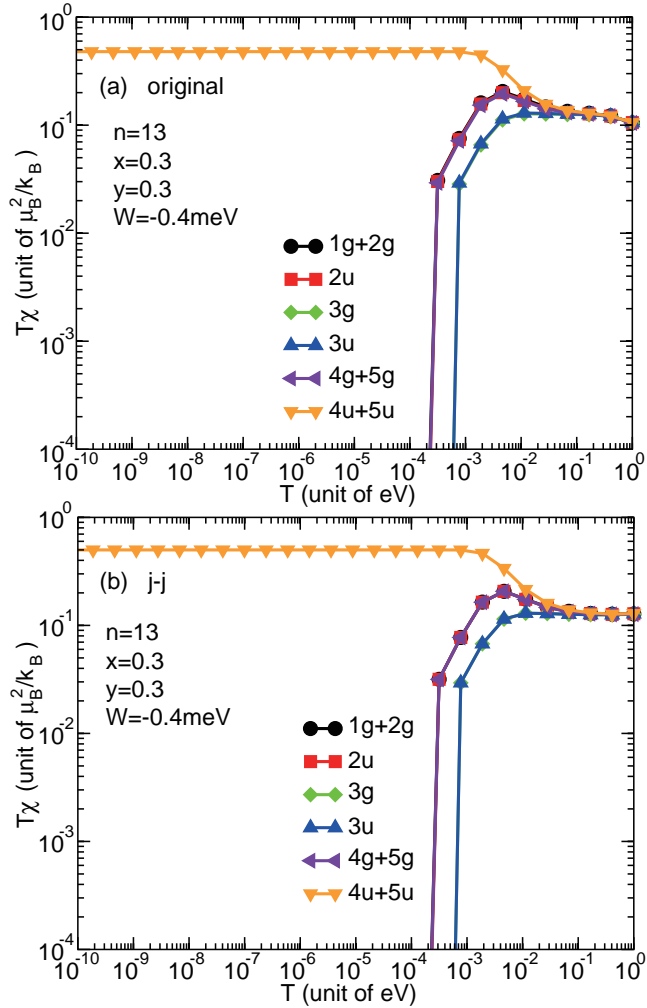}
\caption{(Color online) Multipole susceptibilities of
(a) the original seven-orbital model eq.~(\ref{And-org}) and
(b) the $j$-$j$ coupling model eq.~(\ref{And-jj})
for $n$=13 and $x$=0.3.
}
\end{center}
\end{figure}

In the present Anderson models for filled skutterudites,
we consider only the $a_{\rm u}$ conduction band
composed of $p$ electrons in pnictogens.
As shown in the explanation of the models,
this conduction band is hybridized with $\Gamma_7^-$
state in the $O_{\rm h}$ group.
In the $T_{\rm h}$ group, $\Gamma_6^-$ and $\Gamma_7^-$
are mixed to form two doublets.
Namely, the $\Gamma_7^-$ component of the
CEF ground state for $n$=13 in the $T_{\rm h}$ group
is hybridized with the conduction band, while
the $\Gamma_6^-$ component remains even at low temperatures.
Thus, there occurs residual entropy of $\log 2$
in the present model.
Of course, when we consider further the conduction bands
other than $a_{\rm u}$, the residual entropy should be
finally released at low temperatures.

Let us move on to the numerical results for multipole susceptibility.
In Figs.~4, we show the temperature dependence of eigenvalues $\chi$
of the susceptibility matrix which are classified by the symmetry.
Note that we plot $T\chi$, not $\chi$, which is the Curie constant
for the multipole susceptibility.
In this case, the difference between both models is not so significant
even at high temperatures.
At low temperatures, the magnitude of multipole susceptibility
in Fig.~4(a) is slightly different from that in Fig.~4(b),
but its difference is very small.

\begin{table}
\begin{tabular}{c|c|r|r}
\hline
  rank $k$ & $\gamma$ & Seven-orbital model & $j$-$j$ coupling model\\
\hline
  $1$ &  4u  &  $-0.25091$ & $-0.29008$ \\
\hline
  $3$ &  4u  &  $0.18220$  & $0.18893$ \\
\hline
  $3$ &  5u  &  $0.24455$  & $0.26034$ \\
\hline
  $5$ &  4u (1) &  $0.66679$  & $0.65432$ \\
\hline
  $5$ &  4u (2)  &  $0.11779$ & $0.11847$ \\
\hline
  $5$ &  5u  &  $-0.11770$  & $-0.11974$ \\
\hline
  $7$ &  4u (1)  &  $0.03571$  & $0.03743$ \\
\hline
  $7$ &  4u (2) &  $-0.57753$  & $-0.56282$ \\
\hline
  $7$ &  5u (2) &  $-0.19299$  & $-0.19418$ \\ \hline
\end{tabular}
\caption{Coefficients $p^{(k)}_{\gamma}$
of low-temperature multipole state
eq.~(\ref{multi}) with the largest eigenvalue
in the original seven-orbital model eq.~(\ref{And-org}) and
the $j$-$j$ coupling model eq.~(\ref{And-jj})
for $n$=13 and $x$=0.3.}
\end{table}

In order to confirm the effectiveness of the $j$-$j$ coupling model,
let us turn our attention to the multipole states, not the eigenvalues,
of the multipole susceptibility.
In Table~I, we explicitly list the numbers of the component
$p^{(k)}_{\gamma}$ of
the multipole state at low enough temperatures for both models.
First we note that 4u and 5u are mixed due to the effect of
the $T_{\rm h}$ group and higher-order multipoles are
also included with significant weights.
In general, there is no explicit relation between
admixture and rank in the multipole state.
It is not surprising to obtain significant components of
higher-order multipoles.
When we compare the value of each component,
of course, there exists difference between two models,
but we can conclude in a satisfactory level that
the multipole state of the original seven-orbital Anderson model
is reproduced by the $j$-$j$ coupling model.
Thus, the $j$-$j$ coupling model is useful to analyze the
$f$-electron state with the use of small numbers of
relevant $f$ orbitals.

Next we consider the case of $n$=12.
First we set $x$ as $x$=0.3 which is considered to be
an appropriate value for filled skutterudites ROs$_4$Sb$_{12}$,
even if rare-earth atom R is substituted.
For $n$=12, as observed in Figs.~2, the CEF ground state is
$\Gamma_1^+$ singlet.
In Figs.~5, we show the numerical results of entropy and specific heat
both for the original seven-orbital and the $j$-$j$ coupling models.
Again we see that both panels agrees well with each other,
except for difference in the high-temperature region $T$$\sim$1.
Thus, we reconfirm that the $j$-$j$ coupling model works well
also for the case of $n$=12 with two $f$ holes.

\begin{figure}[t]
\begin{center}
\includegraphics[width=8.5truecm]{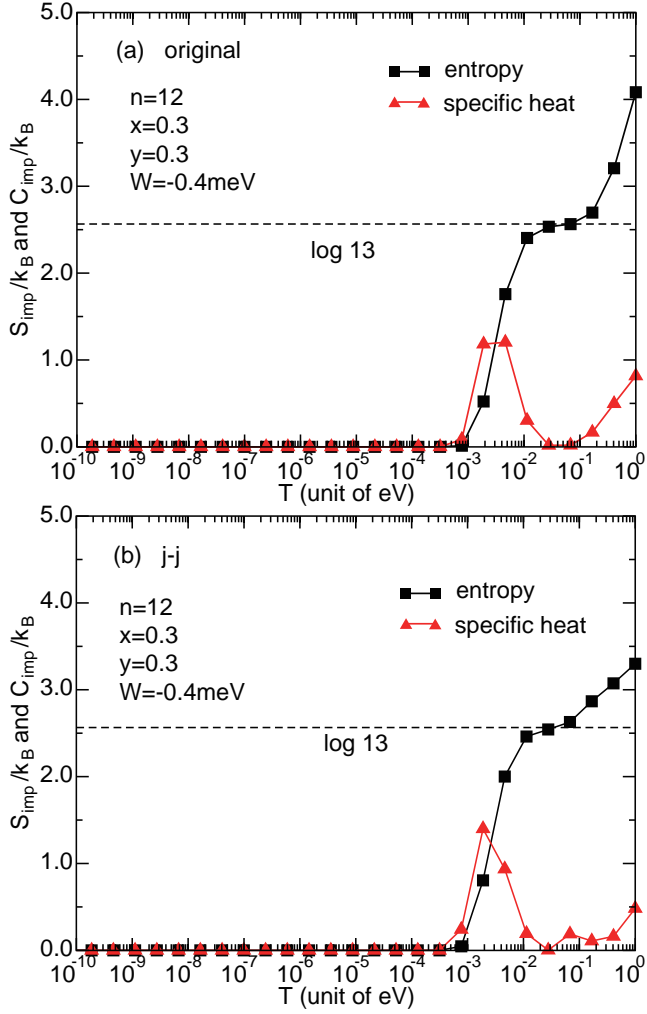}
\caption{(Color online)
Entropy $S_{\rm imp}$ and specific heat $C_{\rm imp}$ of
(a) the original model seven-orbital eq.~(\ref{And-org}) and
(b) the $j$-$j$ coupling model eq.~(\ref{And-jj})
for $n$=12 and $x$=0.3.
}
\end{center}
\end{figure}

For a temperature between 0.01 and 0.1,
we observe a short plateau of $\log 13$ in the entropy,
which is considered to be due to 13-fold degeneracy
of the $J$=6 state.
Then, the entropy of $\log 13$ is released to arrive at
the singlet ground state.
In the singlet ground state, we expect no multipole moment.
In fact, as shown in Figs.~6, multipole susceptibilities
vanish at a temperature at which the specific heat shows a peak
due to the release of entropy $\log 13$.
Note that for the $j$-$j$ coupling model, at high temperatures,
we see significant difference in multipole susceptibilities
between Figs.~6(a) and 6(b).
Probably it is due to the difference in high-energy local states
between the original seven-orbital and the $j$-$j$ coupling models
in the combination with the lack of the renormalization steps.
In any case, when we further make the renormalization process,
we finally obtain the same behavior in multipole susceptibility.

\begin{figure}[t]
\begin{center}
\includegraphics[width=8.5truecm]{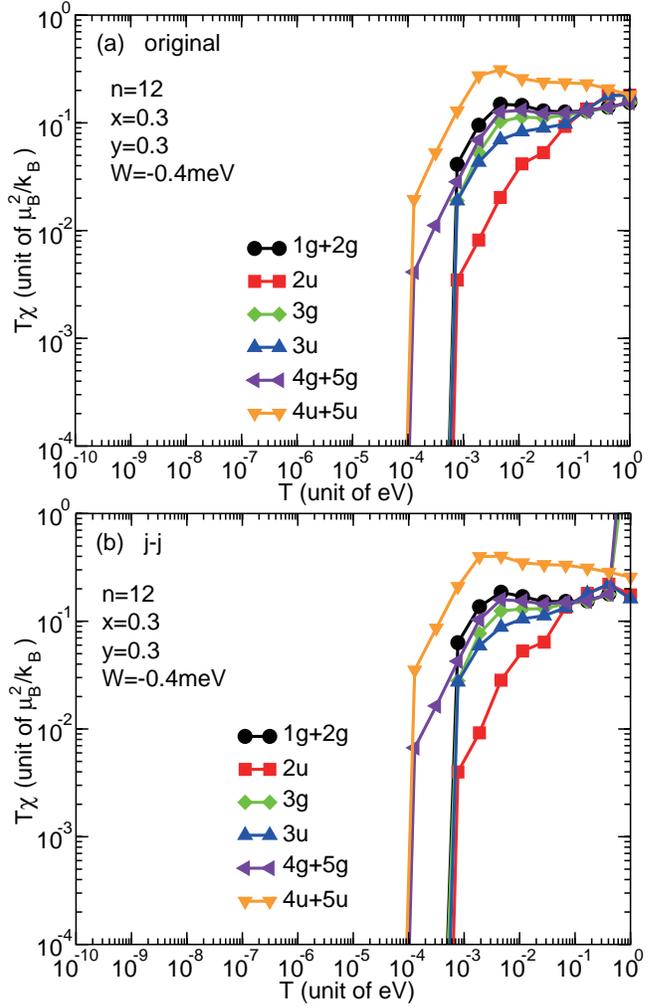}
\caption{(Color online)
Multipole susceptibilities of
(a) the original seven-orbital model eq.~(\ref{And-org}) and
(b) the $j$-$j$ coupling model eq.~(\ref{And-jj})
for $n$=12 and $x$=0.3.
}
\end{center}
\end{figure}

Even at low enough temperatures, no multipole susceptibility is observed
for $n$=12 and $x$=0.3, but it is interesting to consider a possibility
of heavy-electron state in Tm-based filled skutterudites.
Namely, on the basis of the present numerical calculations,
we observe large entropy release such as $\log 13$
at relatively high temperature.
It may be risky to conclude the heavy-electron state only from
the present results, but it seems to be interesting to perform
the measurements of basic bulk properties of Tm-based filled skutterudites,
although it may be difficult to
synthesize actually Tm-based filled skutterudite compounds.

Let us again turn our attention to multipole state for $n$=12.
Here we increase the value of $x$ by assuming that the value of $x$
is controlled experimentally due to the substitution of
transition metal atoms and/or pnictogens.
In Figs.~7 and 8, we show the numerical results for $n$=12 and $x$=1.0.
Note that at $x$=1.0, the local CEF ground state is
$\Gamma_{23}^{+}$ non-Kramers doublet with the first excited state
of $\Gamma_4^+$ triplet, as observed in Figs.~2.

In Fig.~7(a), we show the results of entropy and specific heat.
Except for the high-temperature region,
we again observe that both panels are similar to each other.
In this case, after a short plateau of $\log 13$ around at $T$$\sim$0.1,
we observe the remnant of plateau of $\log 5$
due to 5-fold degeneracy of quasi-quintet
composed of $\Gamma_{23}^{+}$ doublet and $\Gamma_4^+$ triplet.
Then, we arrive at the residual doublet state
composed of a couple of electrons in $\Gamma_{67}^-$
quartet state ($\Gamma_8^-$ in the $O_{\rm h}$ group).
Since in the present model, we consider only the single $a_{\rm u}$
conduction band, which is hybridized with $\Gamma_7^-$ component
of $\Gamma_{5}^-$ doublet state,
$f$ electrons in $\Gamma_{67}^-$ states are considered to be localized.
As mentioned above, in actual materials, the residual $\log 2$ entropy
should be finally released, since there exist other conduction bands
such as $e_{\rm u}$ which hybridize with $\Gamma_{67}^-$ states.

\begin{figure}[t]
\begin{center}
\includegraphics[width=8.5truecm]{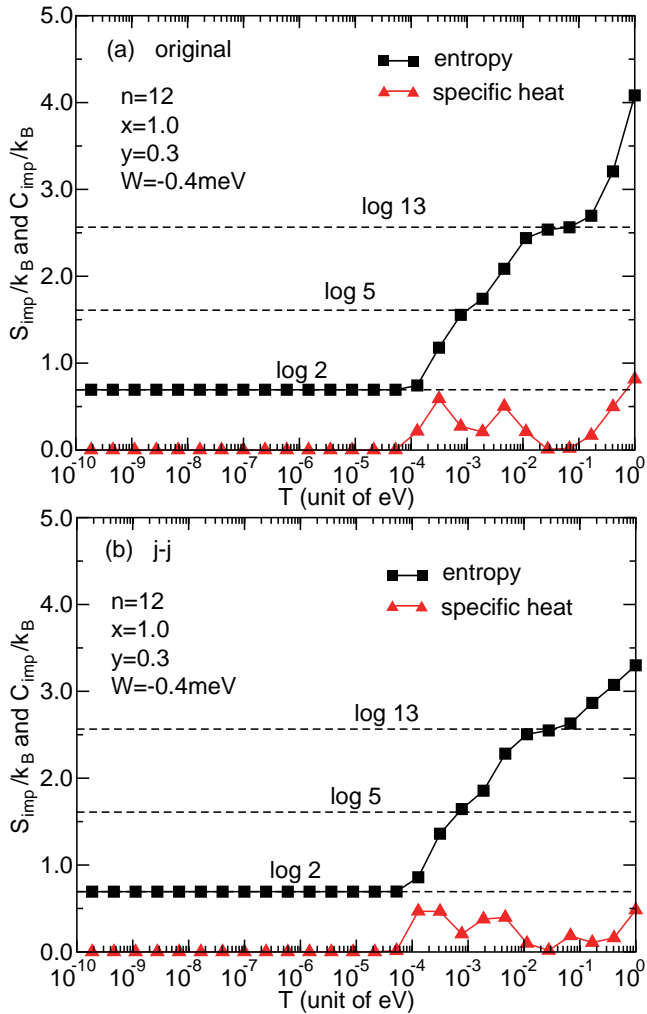}
\caption{(Color online)
Entropy $S_{\rm imp}$ and specific heat $C_{\rm imp}$ of
(a) the original model seven-orbital eq.~(\ref{And-org}) and
(b) the $j$-$j$ coupling model eq.~(\ref{And-jj})
for $n$=12 and $x$=1.0.
}
\end{center}
\end{figure}

Let us explain the results for multipole susceptibility.
It is observed that except for the high-temperature region
larger than $T$=0.1, Figs.~8(a) and 8(b) agree well with each other.
At high temperatures,
significant difference can be found in multipole susceptibilities
between Figs.~8(a) and 8(b),
but it is due to the same reasons as those in Figs.~6.
At low enough temperatures, we find two kinds of residual
multipole states which are expected to be dominant in actual materials,
although ordering type cannot be specified by the present calculations.
The eigenstate with the largest eigenvalue is found to be
characterized by 3g,
while the eigenstate with the second largest eigenvalue
is labelled by 2u.

\begin{figure}[t]
\begin{center}
\includegraphics[width=8.5truecm]{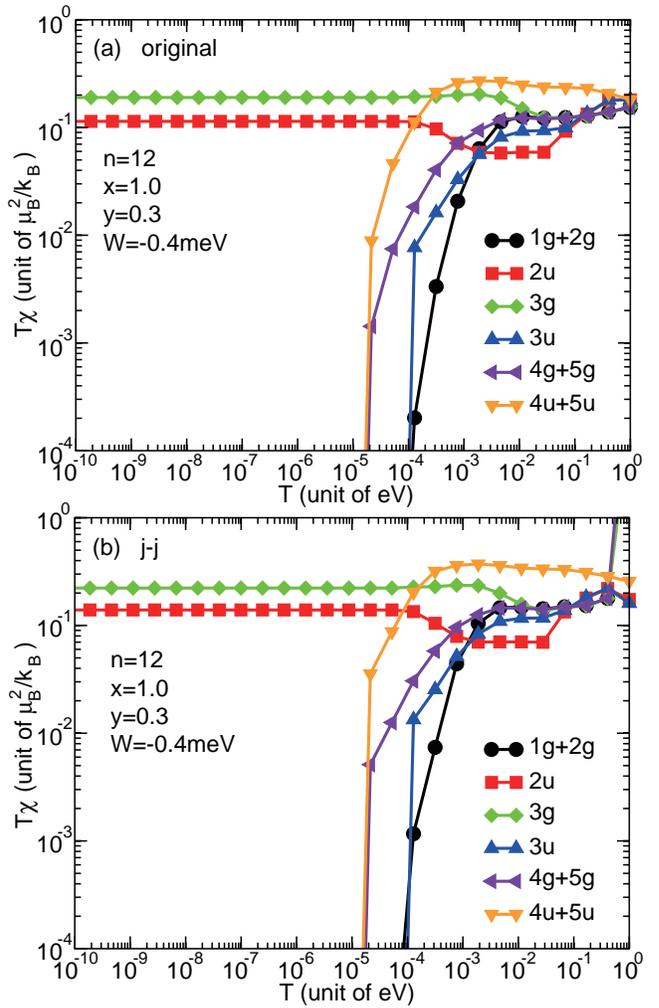}
\caption{(Color online)
Multipole susceptibilities of
(a) the original seven-orbital model eq.~(\ref{And-org}) and
(b) the $j$-$j$ coupling model eq.~(\ref{And-jj})
for $n$=12 and $x$=1.0.
}
\end{center}
\end{figure}

\begin{table}[t]
\begin{tabular}{c|c|r|r}
\hline
  rank $k$ & $\gamma$ & Seven-orbital model & $j$-$j$ coupling model\\
\hline
  $2$ &  3g  &  $-0.70572$ & $-0.69780$ \\
\hline
  $4$ &  3g  &  $-0.01737$  & $-0.00369$ \\
\hline
  $6$ &  3g  &  $0.70828$  & $0.71628$ \\
\hline
\end{tabular}
\caption{Coefficients $p^{(k)}_{\gamma}$
of low-temperature multipole state
eq.~(\ref{multi}) with the largest eigenvalue
in the original seven-orbital model eq.~(\ref{And-org}) and
the $j$-$j$ coupling model eq.~(\ref{And-jj})
for $n$=12 and $x$=1.0.}
\end{table}

\begin{table}[t]
\begin{tabular}{c|c|r|r}
\hline
  rank $k$ & $\gamma$ & Seven-orbital model & $j$-$j$ coupling model\\
\hline
  $3$ &  2u  &  $-0.05988$ & $-0.01126$ \\
\hline
  $7$ &  2u  &  $0.99821$  & $0.99994$ \\
\hline
\end{tabular}
\caption{Coefficients $p^{(k)}_{\gamma}$
of low-temperature multipole state
eq.~(\ref{multi}) with the second largest eigenvalue
in the original seven-orbital model eq.~(\ref{And-org}) and
the $j$-$j$ coupling model eq.~(\ref{And-jj})
for $n$=12 and $x$=1.0.}
\end{table}

In order to examine the multipole state,
in Table II, we show $p_{\gamma}^{(k)}$
of the eigenstate with the largest eigenvalue.
We find small difference in the values between the original seven-orbital
and the $j$-$j$ coupling models,
but it can be concluded that the $j$-$j$ coupling model works well
for the description of the multipole state for $n$=12.
In the multipole state characterized by 3g,
we find two significant components of rank 2 (quadrupole) and rank 6
(tetrahexacontapole), while the rank 4 component (hexadecapole) is
negligibly small.
We are inclined to think that
the present 3g multipole state is expressed by
anti-bonding combination of quadrupole and tetrahexacontapole,
although we cannot prove it analytically at this stage.

In Table III, we show the components for the eigenstate with
the second largest eigenvalue.
This is the multipole characterized by 2u,
which is expected to appear in filled skutterudite structure
due to the localized nature of electrons in $\Gamma_{67}^-$.
\cite{Hotta5,Hotta6}
Interestingly enough, the main component is not octupole
(rank 3) as expected in the case of $n$=5,\cite{Hotta5,Hotta6}
but the rank-7 component (octacosahectapole) becomes dominant.
It may be concluded that pure 2u octacosahectapole occurs
from the numerical results of both the original seven-orbital
and the $j$-$j$ coupling models.
Such a high-rank multipole has been never observed and
a way to detect it experimentally is not known.
However, we expect that exotic ground state including higher-order
multipoles such as rank 6 and 7 is realized
in Tm-based filled skutterudites.

%
%
\section{Discussion and Summary}

In this paper, we have proposed the microscopic model for
Yb- and Tm-based compounds on the basis of the $j$-$j$ coupling scheme.
We have analyzed the impurity Anderson model
in the $j$-$j$ coupling scheme
with the use of a numerical renormalization group technique.
The results have indicated that the $j$-$j$ coupling model works well
for the microscopic description of the multi-$f$-electron state.

Note, however, that in the present paper, we have considered only
the single conduction band in the Anderson model.
Namely, some orbitals are assumed to be localized,
but in actual situations, $f$ electrons in all orbitals should be,
more or less, hybridized with conduction bands.
The present Anderson model is useful to select the candidates of
possible multipoles at low temperatures in an unbiased manner,
although it is not enough to understand low-temperature properties
of actual materials quantitatively.

In order to consider simultaneously the formation of heavy electron state
and the appearance of magnetism and/or superconductivity,
it is necessary to analyze the periodic Anderson model.
It is not difficult to write down the Hamiltonian
for the the multiorbital periodic Anderson model
on the basis of the present prescription of the $j$-$j$ coupling scheme
for Yb and Tm compounds.
As for magnetism, one way is to derive the orbital dependent
RKKY interaction for the determination of the type of multipole ordering.
Such calculations may be performed, although it is difficult
to discuss the competition with Kondo effect.
The actual analysis of the multiorbital periodic Anderson model
in the $j$-$j$ coupling scheme will be an important issue
to be resolved in future.

As for emergence of superconductivity,
first we assume that heavy-electron states are formed.
Then, we consider multiorbital Hubbard-like model for quasi-particles
with the local interaction in the $j$-$j$ coupling scheme,
which is written as
\begin{equation}
 \label{jj-Hubbard}
 \begin{split}
 H & =\sum_{\mib{i},\mib{a},\mu,\nu}
 {\tilde t}_{\mu,\nu}^{\mib{a}}
 f_{\mib{i}\mu}^{\dag} f_{\mib{i}+\mib{a}\nu}
 +\sum_{\mib{i},\mu,\nu}
 {\tilde B}_{\mu,\nu} f_{\mib{i}\mu}^{\dag} f_{\mib{i}\nu} \\
 & +\sum_{\mib{i},\mu,\nu,\mu',\nu'}
 {\tilde I}_{\mu,\nu;\nu',\mu'}
 f_{\mib{i}\mu}^{\dag} f_{\mib{i}\nu}^{\dag}
 f_{\mib{i}\nu'} f_{\mib{i}\mu'},
 \end{split}
\end{equation}
where $f_{\mib{i}\mu}$ is the annihilation operator for $f$ electron
at site $\mib{i}$ with $z$-component $\mu$ of $j$=7/2,
${\tilde t}_{\mu,\nu}^{\mib{a}}$
is $f$-electron effective hopping between $\mu$- and $\nu$-orbitals
along a direction
specified by $\mib{a}$ connecting adjacent two sites,
${\tilde B}$ is given by eq.~(\ref{jj-CEF}),
and ${\tilde I}$ are given by eqs.~(\ref{Eq:Jz6})$-$(\ref{Eq:Jz0}).
Here ${\tilde t}$ is evaluated by the tight-binding approximation
and it is expressed with the use of Slater-Koster integrals,
$(ff\sigma)$, $(ff\pi)$, $(ff\delta)$, and $(ff\phi)$.
\cite{Slater-Koster,Takegahara2}
The model eq.~(\ref{jj-Hubbard}) will be analyzed,
for instance, within a random phase approximation.
Then, it will be necessary to proceed to
the fluctuation-exchange approximation.
In any case, we expect the emergence of exotic superconductivity
in the vicinity of ordered state.
The multipole ordering can be also discussed in the same scheme.
It is one of future directions of the research.

We emphasize that microscopic understanding
of magnetism and superconductivity is one of important issues
in the research field of strongly correlated $f$-electron systems.
The microscopic research will be difficult
on the basis of the $LS$ coupling scheme,
but it is possible with the use of standard field-theoretical techniques
if we exploit the $j$-$j$ coupling model which has been shown
in the present paper.
We note that the applicability of the $j$-$j$ coupling scheme is
wider than one has naively expected from the standard textbook and
it works even for the realistic parameter region
concerning spin-orbit coupling and Coulomb interactions.
We expect that the microscopic research on $f$-electron systems can be
further pushed in future with the use of the $j$-$j$ coupling model.

In summary, we have proposed the prescription to construct
the effective microscopic model for heavy lanthanide systems
such as Yb and Tm compounds on the basis of the $j$-$j$ coupling scheme.
We have numerically analyzed a couple of Anderson models
in which the local interactions at an impurity site are described
by using seven $f$ orbitals and the $j$-$j$ coupling scheme.
We have found that entropy, specific heat, and multipole susceptibilities
are well reproduced by the $j$-$j$ coupling model.
At low enough temperature, the multipole wave function is also well
approximated by the $j$-$j$ coupling scheme in a satisfactorily level
for $n$=13 and 12.

%
%
\section*{Acknowledgement}

The author thanks Y. Aoki and H. Sato for useful discussions
on heavy-fermion compounds.
This work has been supported by a Grant-in-Aid
for Scientific Research on Innovative Areas ``Heavy Electrons''
(No. 20102008) of The Ministry of Education, Culture, Sports,
Science, and Technology, Japan.
The computation in this work has been done
using the facilities of the Supercomputer Center of
Institute for Solid State Physics, University of Tokyo.

%
%


\begin{thebibliography}{99}
\bibitem{note}
Note that the definitions of $E_k$ ($k$=0$\sim$3) for the $j$=7/2 octet
are different from those for the $j$=5/2 sextet
in Refs.~\citen{Hotta1} and \citen{Hotta2}.
\end{thebibliography}
\end{document}